\documentclass[apjl]{emulateapj}
\usepackage{apjfonts}

\newcommand{\etal}{et~al.\ }

\newcommand{\feka}{\hbox{Fe\,K$\alpha$}}

\newcommand{\be}{\begin{equation}}
\newcommand{\ee}{\end{equation}}
\newcommand{\ba}{\begin{eqnarray}}
\newcommand{\ea}{\end{eqnarray}}

\newcommand{\chandra}{{\emph{Chandra}}}
\newcommand{\cxo}{{\emph{Chandra X-ray Observatory}}}

\newcommand{\hst}{\emph{HST}}

\newcommand{\simgt}{\lower 2pt \hbox{$\, \buildrel {\scriptstyle >}\over {\scriptstyle\sim}\,$}}
\newcommand{\simlt}{\lower 2pt \hbox{$\, \buildrel {\scriptstyle <}\over {\scriptstyle\sim}\,$}}
\newcommand{\ls}{\lower 2pt \hbox{$\;\scriptscriptstyle \buildrel<\over\sim\;$}}
\newcommand{\gs}{\lower 2pt \hbox{$\;\scriptscriptstyle \buildrel>\over\sim\;$}}
\newcommand{\sarc}{$^{\prime\prime}\!\!.$}

\begin{document}

\def\arcsec{$^{\prime\prime}$}
\def\arcmin{$^{\prime}$}
\def\degr{$^{\circ}$}

\title{Discovery of Energy Dependent X-ray Microlensing in Q2237+0305}

\author{Bin Chen\altaffilmark{1}, Xinyu Dai\altaffilmark{1}, C. S. Kochanek\altaffilmark{2}, George Chartas\altaffilmark{3}, Jeffrey A. Blackburne\altaffilmark{2}, Szymon Koz{\l}owski\altaffilmark{4}} 

\altaffiltext{1}{Homer L. Dodge Department of Physics and Astronomy,
University of Oklahoma, Norman, OK 73019, USA}

\altaffiltext{2}{Department of Astronomy, The Ohio State University, Columbus, OH 43210, USA}

\altaffiltext{3}{Department of Physics and Astronomy, College of Charleston, SC 29424, USA}

\altaffiltext{4}{Warsaw University Observatory, Al.\ Ujazdowskie 4, 00-478 Warszawa, Poland}

\begin{abstract}
    We present our long term \emph{Chandra} X-ray monitoring data for the gravitationally lensed quasar Q2237+0305 with 20 epochs spanning 10 years.  We easily detect microlensing variability between the images in the full (0.2--8 keV), soft (0.2--2 keV), and hard (2--8 keV) bands at very high confidence.  We also detect, for the first time, chromatic microlensing differences between the soft and hard X-ray bands.  The hard X-ray band is more strongly microlensed than the soft band, suggesting that the corona above the accretion disk thought to generate the X-rays has a non-uniform electron distribution, in which the hotter and more energetic electrons occupy more compact regions surrounding the black holes.
    Both the hard and soft X-ray bands are more strongly microlensed than the optical (restframe UV) emission, indicating that the X-ray emission is more compact than the optical, confirming the microlensing results from other lenses.
\end{abstract}


\section{Introduction}

Quasar microlensing is an important new tool for studies of the structure of quasar accretion disks based on the dependence of the microlensing magnification on the source size (e.g., Wambsganss 2006; Kochanek et al.\ 2007).
Microlensing was first detected by Irwin et al.\ (1989) in Q2237+0305 (Huchra et al.\ 1985).
Afterwards, microlensing effects have been frequently reported in this system in different energy bands (e.g., Wozniak et al.\ 2000; Dai et al.\ 2003; Goicoechea et al.\ 2003; Vakulik et al.\ 2004; Anguita et al.\ 2008; Eigenbrod et al.\ 2008; Mosquera et al.\ 2009; O'Dowd et al.\ 2010).  The quasar microlensing technique was first used to crudely determine emission sizes relative to the Einstein ring size (e.g, Wambsganss et al.\ 1990; Rauch \& Blandford 1991; Falco et al.\ 1996; Mediavilla et al.\ 1998; Agol et al.\ 2000).  However, since the magnification diverges on the caustics, the technique can be used to resolve emission regions arbitrarily smaller than the Einstein ring size, which allowed crude measurements of the X-ray continuum and \feka\ emission regions (e.g., Chartas et al.\ 2002, 2007; Dai et al.\ 2003; Ota et al.\ 2006; Blackburne et al.\ 2006; Pooley et al.\ 2006).
Quasar microlensing models has been developed by different groups to interpret the microlensing results, where a major breakthrough occurred in Kochanek (2004) using a Bayesian Monte Carlo model. 
This approach and its derivatives have successfully reproduced microlensing light curves in the optical and X-ray bands, determined the emission sizes quantitatively (Morgan et al.\ 2008; Chartas et al.\ 2009; Dai et al.\ 2010; Mediaville et al.\ 2011), the stellar/dark matter fractions (e.g., Mediavilla et al.\ 2009; Poindexter \& Kochanek 2010a; Bate et al.\ 2011), and even the inclination angles of accretion disks (Poindexter \& Kochanek 2010b).

The X-ray continuum emission of quasars is generally believed to be created by the un-saturated inverse Compton scattering emission between soft UV photons from the disk and hot electrons in the corona, leading to the observed power-law X-ray spectra (e.g., Reynolds \& Nowak 2003). 
Thus, the size of the X-ray emission traces the extent of the corona.  Because of its shorter variability time scales, the X-ray continuum emission region is believed to be smaller than the optical emission region.  However, the simple variability argument is not conclusive as the variability could be dominated by localized flares rather than disk-scale variability.  Recent microlensing results, however, have conclusively shown that the X-ray emission regions are an order of magnitude smaller than the (observed frame) optical emission regions (e.g., Morgan et al.\ 2008; Chartas et al.\ 2009; Dai et al.\ 2010; Blackburne et al.\ 2011), excluding models where the X-ray coronae covers large portions of the accretion disks.  

Besides the continuum emission, there are additional X-ray components such as the reflection hump, metal emission lines, soft X-ray excess, and jet emission, which may also contribute to the X-ray spectra of quasars.  Quasar microlensing is one of the few tools that can determine the physical size of these different components, and microlensing effects have been detected in the \feka\ emission line (e.g., Chartas et al.\ 2002, 2007; Dai et al.\ 2003; Ota et al.\ 2006).  In addition, if the hot electron distribution is not uniform, so we should observe chromatic, energy dependent microlensing of the X-ray continuum.  In this letter, we report results of our long term \emph{Chandra} X-ray monitoring of Q2237+0305, and the first detection of chromatic microlensing of the X-ray continuum.

\section{Observations And Data Reduction}

In various combinations, we have now observed Q2237+0305 with ACIS (Garmire et al. 2002) on board the \cxo\ (Weisskopf et al.\ 2002) for 20 epochs between September 2000 and November 2010 (Table~\ref{tab:2237count}).   
We reprocessed all these data using the CIAO 4.3 software tools 
by removing the pixel randomization and applying a sub-pixel algorithm for event positions (Tsunemi et al.\ 2001).
We filtered all events using the standard ASCA grades of 0, 2, 3, 4  and 6 with energies between 0.2 and 8 keV (observer's frame), and further separated the events into soft (0.2--2.0 keV) and hard (2.0--8.0 keV) bands for our subsequent analyses. 
For each band,  we extracted the total counts from all 4 images within 3\arcsec\ radius circular regions and the background counts from concentric annular regions with inner/outer radii of 10\arcsec$\>$ and 20\arcsec, respectively. 
We then used the PSF fitting method to model the relative count rates between the images, fixing the relative image positions to the \hst\ positions\footnote{http://cfa-www.harvard.edu/glensdata/}.
We binned the X-ray images in 0\sarc0984, and fit these binned images using \verb+Sherpa+ to minimize the Cash C-statistic between the observed and model images.  
The final count rates were normalized by partitioning the background-subtracted total source counts using the (unnormalized) relative count rates for each image obtained from the PSF fits.    
We also obtained a stacked image of Q2237+0305 (Fig.~1) based on the image positions obtained from the PSF fitting results.

\begin{figure}
	\epsscale{1.0}
        \plotone{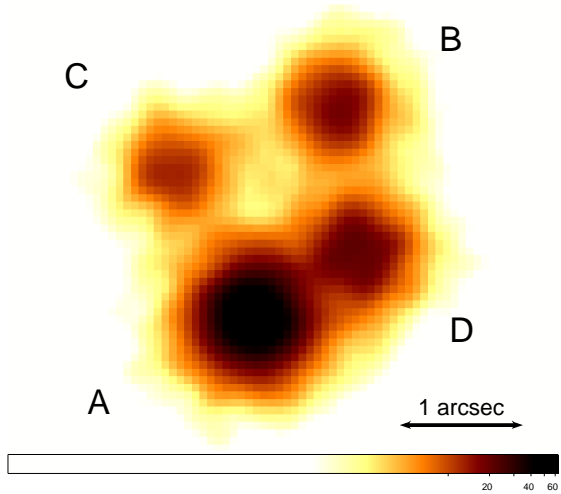}
	\caption{Combined X-ray image of Q2237+0305 from 20 epochs observed between 2000 and 2010. The total integration time is 292~ks. \label{fig:merge}}
\end{figure}

We extracted spectra of the individual images in 0\sarc5 radius circles centered on the PSF fitting positions for each observation.
We obtained combined spectra for each image by stacking the spectra and the corresponding \verb+rmf+ and \verb+arf+ files weighted by their exposure times, and analyzed the combined spectra using XSPEC V12.  
We fit the spectra of the individual (stacked) images using a simple power-law modified by Galactic absorption with $N_H=5.5\times 10^{20}\> {\rm cm}^{-2}$ (Dickey \& Lockman 1990) and neutral absorption at the lens ($z_l=0.0395$) for the 0.4--8.0 keV band. 
We detect differential absorption between the images at the lens redshift. 
After obtaining the best fits, we calculated the ratios of the photon count rates between the absorbed and unabsorbed models for the full, soft, and hard bands, respectively. 
These mean ratios were then applied to the count rates of individual epochs and images to estimate the absorption corrected photon count rates.  The absorption corrected count rates are presented in Table~\ref{tab:2237count}.  
Zimmer et al.\ (2011) computed the photon count rates of Q2237+0305 in the 0.5--10 keV band for 10 \emph{Chandra} observations between 2006 and 2007; however, the authors did not correct for the absorption effects or divide them into energy bands.  
To double check our analyses, we also measured the absorbed count rates for these observations in the 0.5--10 keV band, and our results are consistent.

\section{Analysis}

\begin{figure*}
    \includegraphics[width=1.0\textwidth,height=0.8\textheight]{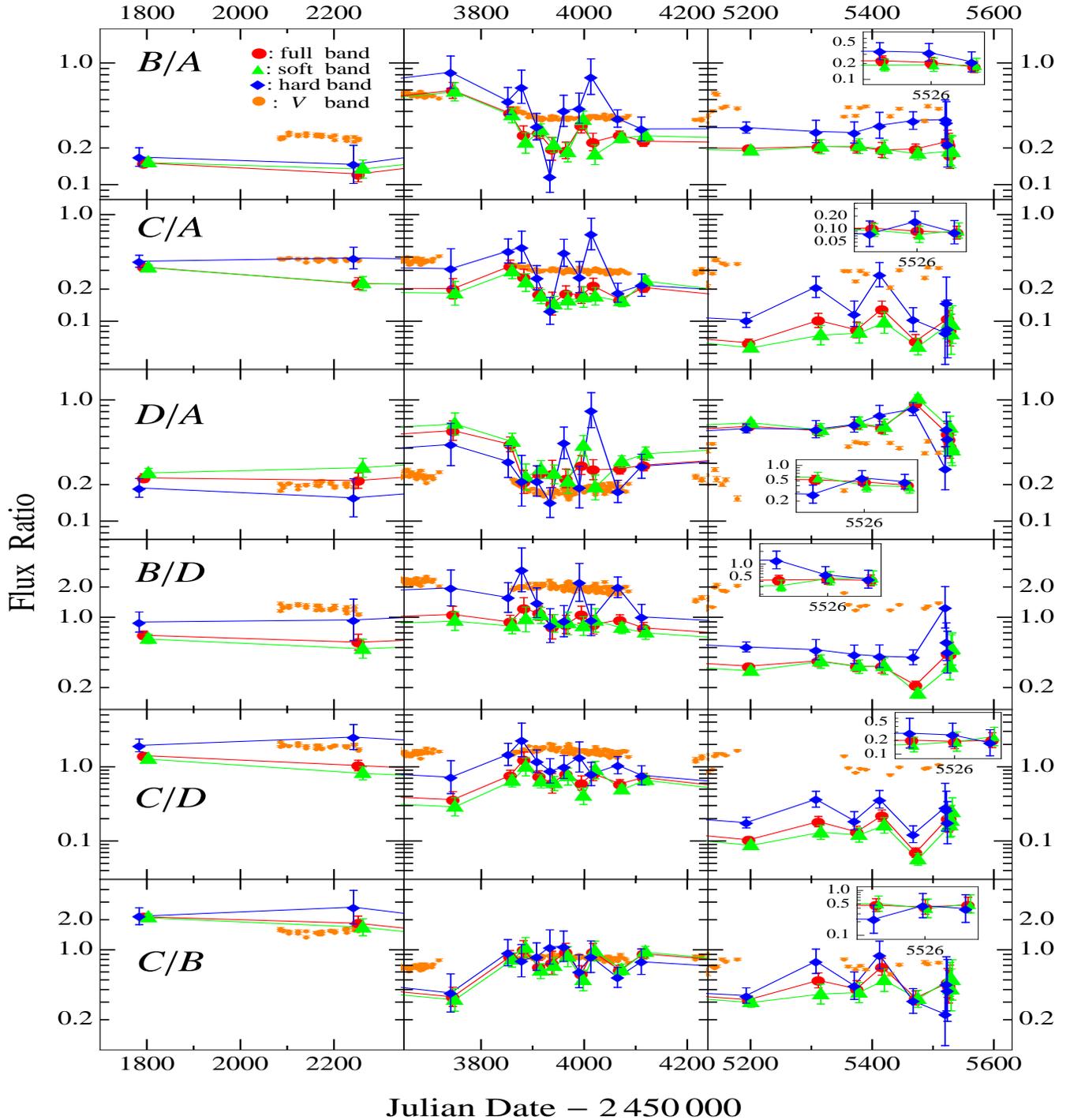}
\caption{Chromatic X-ray microlensing light curves of Q2237+0305 in the full X-ray band (2.0--8.0 keV, red circles), soft X-ray band (0.2--2 keV, green triangles), hard X-ray band (2--8 keV, blue diamonds), and OGLE $V$ band (Wozniak et al.\ 2000; Udalski et al.\ 2006) and our SMARTS data (orange circles).  Insets in the third column are blow-ups of the last X-ray data points (three observations made within 4 days).  We slightly offset the observation dates for the soft and hard bands to make the difference more visible. \label{fig:microlensing}}
\end{figure*}

Fig.~\ref{fig:microlensing} shows the time evolution of the flux ratios between various images. 
Since the time delays between different images of Q2237+0305 are short, less than 1 day, from both theoretical and observational constraints (e.g., Schmidt et al.\ 1998; Dai et al. 2003), the changes in the flux ratios are essentially unaffected by intrinsic variability, and thus must be due to microlensing by stars in the foreground lensing galaxy.  
 
\begin{deluxetable}{ccccrrr}
\tabletypesize{\scriptsize}
\setlength{\tabcolsep}{0.015in} 
\tablecolumns{7}
\tablecaption{\chandra\ Count Rates of Q2237+0305 Corrected for Absorption\label{tab:2237count}}
\tablehead{
\colhead{Ob-}&
\colhead{Julian} & 
\colhead{Expo.} &
\colhead{Im-}& 
\colhead{$\rm{CR_{full} }$ }& 
\colhead{$\rm{CR_{soft}}$}&
\colhead{$\rm{CR_{hard}}$}\\
\colhead{sid}&
\colhead{{\rm Date}}&
\colhead{{\rm Time}}&
\colhead{age}&
\colhead{$(\times10^{-3}$} & 
\colhead{$(\times10^{-3}$} & 
\colhead{$(\times10^{-3}$} \\ 
\colhead{}&
\colhead{}&
\colhead{(s)}&
\colhead{}&
\colhead{s$^{-1}$)}&
\colhead{s$^{-1}$)}&
\colhead{s$^{-1}$)}
}
\startdata
   $   431$   &    $   51794$    &     $   30287$    &   A   &     $ 114.6{\pm   3.7}$   &       $ 112.8{\pm   4.0}$   &      $   9.6{\pm   0.7}$ \\  
   $   431$   &    $   51794$    &     $   30287$    &   B   &     $  17.3{\pm   1.1}$   &       $  17.1{\pm   1.2}$   &      $   1.6{\pm   0.3}$ \\  
   $   431$   &    $   51794$    &     $   30287$    &   C   &     $  36.9{\pm   1.7}$   &       $  35.8{\pm   1.8}$   &      $   3.5{\pm   0.4}$ \\  
   $   431$   &    $   51794$    &     $   30287$    &   D   &     $  26.1{\pm   1.8}$   &       $  28.2{\pm   2.2}$   &      $   1.8{\pm   0.3}$ \\  
\tableline
   $  1632$   &    $   52252$    &     $    9535$    &   A   &     $  93.4{\pm   7.0}$   &       $  85.8{\pm   7.3}$   &      $   8.5{\pm   1.1}$ \\  
   $  1632$   &    $   52252$    &     $    9535$    &   B   &     $  11.4{\pm   1.5}$   &       $  11.5{\pm   1.7}$   &      $   1.2{\pm   0.4}$ \\  
   $  1632$   &    $   52252$    &     $    9535$    &   C   &     $  21.0{\pm   2.2}$   &       $  19.3{\pm   2.4}$   &      $   3.3{\pm   0.6}$ \\  
   $  1632$   &    $   52252$    &     $    9535$    &   D   &     $  20.2{\pm   2.6}$   &       $  23.7{\pm   3.8}$   &      $   1.3{\pm   0.4}$ \\  
\tableline
   $  6831$   &    $   53745$    &     $    7264$    &   A   &     $  38.2{\pm   4.0}$   &       $  35.2{\pm   4.2}$   &      $   4.0{\pm   0.8}$ \\  
   $  6831$   &    $   53745$    &     $    7264$    &   B   &     $  22.7{\pm   2.5}$   &       $  20.4{\pm   2.5}$   &      $   3.4{\pm   0.8}$ \\  
   $  6831$   &    $   53745$    &     $    7264$    &   C   &     $   7.7{\pm   1.5}$   &       $   6.4{\pm   1.5}$   &      $   1.3{\pm   0.5}$ \\  
   $  6831$   &    $   53745$    &     $    7264$    &   D   &     $  21.3{\pm   3.2}$   &       $  22.3{\pm   3.8}$   &      $   1.7{\pm   0.6}$ \\  
\tableline
   $  6832$   &    $   53856$    &     $    7936$    &   A   &     $  75.4{\pm   5.4}$   &       $  73.7{\pm   6.1}$   &      $   7.4{\pm   1.1}$ \\  
   $  6832$   &    $   53856$    &     $    7936$    &   B   &     $  29.1{\pm   2.8}$   &       $  27.2{\pm   2.9}$   &      $   3.6{\pm   0.8}$ \\  
   $  6832$   &    $   53856$    &     $    7936$    &   C   &     $  24.8{\pm   2.7}$   &       $  21.5{\pm   2.7}$   &      $   3.4{\pm   0.7}$ \\  
   $  6832$   &    $   53856$    &     $    7936$    &   D   &     $  32.4{\pm   3.9}$   &       $  33.3{\pm   4.6}$   &      $   2.3{\pm   0.6}$   
\enddata
\tablecomments{
Table~1.\ is published in its entirety in the electronic edition, and a portion is shown here for guidance regarding its form and content.
The full, soft, and hard bands are the observed energy bands between 0.2--8, 0.2--2, and 2--8 keV, respectively. 
}
\end{deluxetable}

\begin{figure}
	\epsscale{1.0}
	\plotone{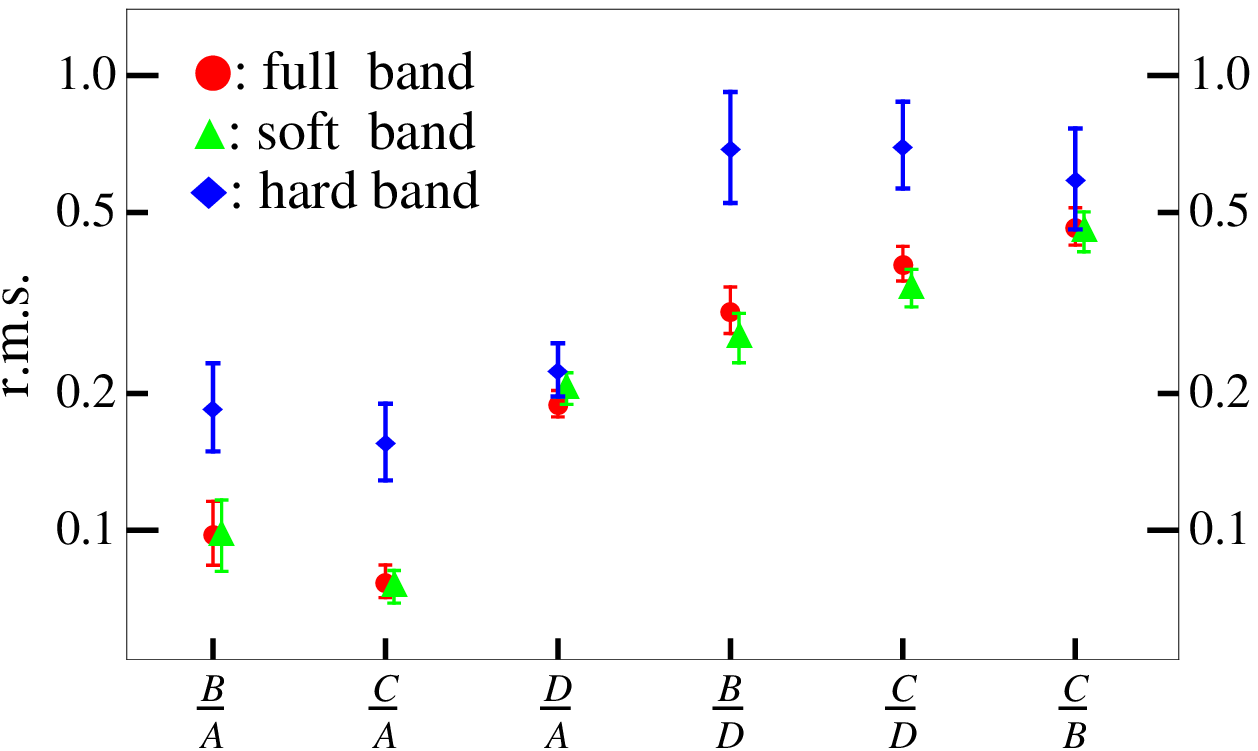}
	\caption{X-ray microlening variability amplitudes in the full, soft and hard bands. 
        \label{fig:Variability}}
\end{figure}

We first fit the full, soft, and hard band light curves in Fig~\ref{fig:microlensing} to constants to test the significance of the microlensing signal.  For the full band, the smallest $\chi^2$ for no microlensing is for the case B/A, with a $\chi^2=97.2$ for 19 degrees of freedom (dof) and a null hypothesis probability of $NP=2\times 10^{-12},$ while the largest is for the case of C/D with $\chi^2=396.8$ and $NP=2\times 10^{-72}.$ For the soft band, the smallest $\chi^2=70.2$ is for B/A with $NP=8\times 10^{-8}$ and the largest $\chi^2=312.0$ is for C/A with $NP=7\times 10^{-55}.$ For the hard band, the smallest is $\chi^2=44.35$ for B/D with $NP=0.00084$, and the largest $\chi^2=147.95$ is for D/A with $NP=5\times 10^{-22}.$
The null hypothesis is ruled out for all 6 image pairs for the full, soft, and hard bands. X-ray microlensing is detected with very high confidence. 

 Next, we compare the hard and soft bands to test for chromatic microlensing. 
Assuming no chromatic microlensing, we compute the $\chi^2$ for the hard and soft bands showing the same flux ratio evolution 
for three ways of combining the images into pairs, i.e.,  B/A plus C/D,  C/A plus B/D, and D/A plus C/B. 
The $\chi^2$ value for B/A+C/D is $\chi^2=89.8$ for $40$ dof with a null hypothesis probability of $NP=1.1\times 10^{-5}.$ For C/A+B/D, we find $\chi^2=91.0$ with  $NP=7.6\times 10^{-6}$. For the D/A+C/B pairs, we find $\chi^2=50.5$ with  $NP=0.12.$ These results indicate that there is significant chromatic X-ray microlensing in Q2237+0305.  Fig.~\ref{fig:microlensing} also shows that the hard band microlensing variability of Q2237+0305 has a larger amplitude than that in the soft band. 
To compare the amplitudes of variability in the soft and hard bands, we calculate the r.m.s.\ variability for each image pair and show the results in Fig~\ref{fig:Variability}. 
The soft and hard microlensing variability amplitudes are different at the 99\% confidence for the B/A plus C/D pairs, and at the 99.7\% confidence for the C/A plus B/D pairs.
 In the last cases, D/A and C/B, the difference in the variability amplitudes are not well resolved.
 
\section{Discussion}

As in previous studies, we detect significant microlensing in Q2237+0305 in the full X-ray band, where the flux ratio changes are dramatically different from those in the optical microlensing light curves, as shown in Fig.~2.
These results confirm the earlier studies in other gravitational lenses that the X-ray continuum emission regions are more compact than the optical emission regions (Morgan et al.\ 2008; Chartas et al.\ 2009; Dai et al.\ 2010).
 By dividing the X-ray energy band (0.2--8.0 keV) into soft and hard bands, we detect for the first time chromatic microlensing of the X-ray continuum in the sense the hard band is more strongly microlensed than the soft. 
For our radio-quiet quasar at $z_s = 1.69$, the observed 0.2--8 keV band probes the rest frame 0.5--22 keV band.
Our spectral analysis does not reveal a reflection or a soft X-ray excess component.
The metal emission lines, e.g., \feka, will fall in this band; however, their equivalent widths are usually small enough such that the broad band flux is little affected.
Therefore, the chromatic X-ray microlensing occurs in the X-ray continuum emission in Q2237+0305.
We note that if additional X-ray emission components exist in the quasar spectra, the interpretation of chromatic X-ray microlensing could be more complicated.

  If the X-ray continuum emission was generated through the inverse Compton scattering of the soft UV photons from the accretion disk  by hot electrons from the corona (Reynolds \& Nowak  2003), our results indicate that the distribution of hot electrons in the corona is not uniform, and that the hot electrons are more compact than the cool electrons.  
Thus, even if the observed X-ray continua emission can be modeled as simple power-laws, our results indicate that the real emission mechanism has spatial structures. 
The evidence of structure in the ``corona'' favors complicated geometrical configuration such as the light bending (Fabian et al.\ 2005) or aborted jet models (Ghisellini et al.\ 2004).
Previous microlensing analyses have constrained the size of the X-ray continuum to about $\sim10 r_g$  (e.g., Morgan et al.\ 2008; Chartas et al. 2009; Dai et al. 2010), and our results in this paper suggest even smaller hard emission regions tracking the event horizon of the black holes.
  A detailed microlensing model of the microlensed X-ray emission from Q2237+0305 will be presented in a forthcoming paper.

\acknowledgements 
We acknowledge financial support by NASA/SAO grants GO6-7093X, GO0-11121A/B/C, GO1-12139A/B/C and NSF grants AST-0708082 and AST-1009756.


\begin{thebibliography}{}
\bibitem[Agol, Jones, \& Blaes(2000)]{a00} Agol, E., Jones, B., \& Blaes, O. 2000, \apj, 545, 657
\bibitem[Anguita \etal(2008)]{an08}Anguita, T., Schmidt, R.W., Turner, E.L., et al.\ 2008, A\&A, 480, 327
\bibitem[Bate et al.(2011)]{2011ApJ...731...71B} Bate, N.~F., Floyd, D.~J.~E., Webster, R.~L., \& Wyithe, J.~S.~B.\ 2011, \apj, 731, 71 
\bibitem[Blackburne et al.(2006)]{bl06}Blackburne, J.A., Pooley, D., \& Rappaport, S., 2006, ApJ, 640, 569
\bibitem[Blackburne et al.(2011)]{2011ApJ...729...34B} Blackburne, J.~A., Pooley, D., Rappaport, S., \& Schechter, P.~L.\ 2011, \apj, 729, 34 
\bibitem[Chartas \etal (2002)]{c02} Chartas, G., Agol, E., Eracleous, M., et al.\ 2002, \apj, 568, 509
\bibitem[Chartas et al.(2007)]{2007ApJ...661..678C} Chartas, G., Eracleous, M., Dai, X., Agol, E., \& Gallagher, S.\ 2007, \apj, 661, 678 
\bibitem[Chartas et al.\ (2009)]{ch09}Chartas, G., Kochanek, C.S., Dai, X., Poindexter, S., \& Garmire, G., 2009, ApJ, 693, 174
\bibitem[Dai et al.\ (2003)]{da03}Dai, X., Chartas, G., Agol, E., Bautz, M.W., \& Garmire, G.P., 2003, ApJ, 589, 100
\bibitem[Dai et al.(2010)]{2010ApJ...709..278D} Dai, X., Kochanek, C.~S., Chartas, G., et al.\ 2010, \apj, 709, 278 
\bibitem[Dickey \& Lockman (1990)]{d90} Dickey, J. M., \& Lockman F. J. 1990, ARA\&A 28, 215
\bibitem[Eigenbrod et al.\ (2008)]{ei08}Eigenbrod, A., Courbin, F., Meylan, G., et al.\ 2008, A\&A, 490, 933
\bibitem[Fabian et al.\ (2005)]{fa05}Fabian, A.C., Miniutti, G., Iwasawa, K., \& Ross, R.R., 2005, MNRAS, 361, 795
\bibitem[Falco \etal(1996)]{f96} Falco, E. E., Leh$\acute{\rm{a}}$r, J., Perley, R. A., Wambsganss, J., \& Gorenstein, M. V. 1996, \aj, 112, 897
\bibitem[Garmire \etal (2002)]{g02} Garmire, G. P., Bautz, M. W., Nousek, J. A., \& Ricker, G. R. 2002, SPIE, 4851
\bibitem[Ghisellini et al.(2004)]{2004A&A...413..535G} Ghisellini, G., Haardt, F., \& Matt, G.\ 2004, \aap, 413, 535 
\bibitem[Goicoechea et al.(2003)]{go03}Goicoechea, L.J., Alcaide, D., Mediavilla, E., Mun\~oz, J.A., 2003, A\&A, 397, 517
\bibitem[Huchra \etal(1985)]{h85} Huchra, J., Gorenstein, M., Kent, S., et al.\ 1985, \aj, 90, 691
\bibitem[Irwin \etal(1989)]{i89} Irwin, M. J., Webster, R. L., Hewett, P. C., Corrigan, R. T., \& Jedrzejewski, R. I. 1989, \aj, 98, 1989
\bibitem[Kochanek (2004)]{ko04}Kochanek C.S., 2004, ApJ, 605, 58
\bibitem[Kochanek et al.(2007)]{2007ASPC..371...43K} Kochanek, C.~S., Dai, X., Morgan, C., et al.\ 2007, Statistical Challenges in Modern Astronomy IV, 371, 43 [astro-ph/0609112] 
\bibitem[Mediavilla \etal (1998)]{m98} Mediavilla, E., Arribas, S., del Burgo, C., et al.\ 1998, \apj, 503, L27
\bibitem[Mediavilla et al.(2009)]{2009ApJ...706.1451M} Mediavilla, E., et al.\ 2009, \apj, 706, 1451 
\bibitem[Mediavilla et al.(2011)]{2011ApJ...730...16M} Mediavilla, E., et al.\ 2011, \apj, 730, 16 
\bibitem[Morgan et al.(2008)]{2008ApJ...689..755M} Morgan, C.~W., Kochanek, C.~S., Dai, X., Morgan, N.~D., \& Falco, E.~E.\ 2008, \apj, 689, 755 
\bibitem[Mosquera et al.(2009)]{2009ApJ...691.1292M} Mosquera, A.~M., Mu{\~n}oz, J.~A., \& Mediavilla, E.\ 2009, \apj, 691, 1292 
\bibitem[O'Dowd et al.(2010)]{2010arXiv1012.3480O} O'Dowd, M., Bate, N.~F., Webster, R.~L., Wayth, R., \& Labrie, K.\ 2010, arXiv:1012.3480 
\bibitem[Ota et al.(2006)]{2006ApJ...647..215O} Ota, N., et al.\ 2006, \apj, 647, 215 
\bibitem[Pooley et al.(2006)]{2006ApJ...648...67P} Pooley, D., Blackburne, J.~A., Rappaport, S., Schechter, P.~L., \& Fong, W.-F.\ 2006, \apj, 648, 67 
\bibitem[Poindexter \& Kochanek(2010)]{2010ApJ...712..658P} Poindexter, S., \& Kochanek, C.~S.\ 2010a, \apj, 712, 658 
\bibitem[Poindexter \& Kochanek(2010)]{2010ApJ...712..668P} Poindexter, S., \& Kochanek, C.~S.\ 2010b, \apj, 712, 668 
\bibitem[Rauch \& Blandford(1991)]{1991ApJ...381L..39R} Rauch, K.~P., \& Blandford, R.~D.\ 1991, \apjl, 381, L39 
\bibitem[Reynolds \& Nowak (2003)]{rn03}Reynolds, C.S., \& Nowak, M.A., 2003, PhR, 377, 389
\bibitem[Schmidt et al.(1998)]{1998MNRAS.295..488S} Schmidt, R., Webster, R.~L., \& Lewis, G.~F.\ 1998, \mnras, 295, 488 
\bibitem[Tsunemi \etal(2001)]{t01} Tsunemi, H., Mori, K., Miyata, E., et al.\ 2001, \apj, 554, 496
\bibitem[Udalski et al.(2006)]{u06} Udalski, A., et al., \ 2006, Acta Astron., 56, 293
\bibitem[Vakulik et al.(2004)]{2004A&A...420..447V} Vakulik, V.~G., et al.\ 2004, \aap, 420, 447 
\bibitem[Wambsganss et al.(1990)]{1990ApJ...358L..33W} Wambsganss, J., Paczynski, B., \& Schneider, P.\ 1990, \apjl, 358, L33 
\bibitem[Wambsganss(2006)]{2006glsw.conf..453W} Wambsganss, J.\ 2006, Saas-Fee Advanced Course 33: Gravitational Lensing: Strong, Weak and Micro, 453 
\bibitem[Weisskopf \etal(2002)]{we02} Weisskopf, M. C., Brinkman, B., Canizares,
 C., et al.\ 2002, \pasp, 114, 1 
\bibitem[Wo$\acute{\rm{z}}$niak \etal (2000a)]{w00a} Wo$\acute{\rm{z}}$niak, P. R., Alard, C., et al.\ 2000, \apj, 529, 88
\bibitem[Zimmer \etal (2011)]{z11} Zimmer, F., Schmidt, R. W., \& Wambsganss, J., 2011, \mnras, 413, 1099
\end{thebibliography}
\end{document}